\documentstyle[preprint,prl,aps,epsf]{revtex}
\begin{document}
\preprint{\vbox{\hbox {May 1998} \hbox{IFP-761-UNC} } }

\title{\bf Cosmic Background Radiation Temperature Anisotropy: 
Position of First Doppler Peak}
\author{\bf Paul H. Frampton, Y.Jack Ng and Ryan Rohm}
\address{Department of Physics and Astronomy,}
\address{University of North Carolina, Chapel Hill, NC  27599-3255}
\maketitle
\begin{abstract}
The purpose of the Cosmic Background Radiation
(CBR) experiments is to measure the temperature
anisotropy via the autocorrelation function.
The partial wave $l_1$ corresponding to the first Doppler
peak caused by baryon-photon oscillations at the surface of
last scattering depends on the present density $\Omega_0$
and the cosmological constant contribution $\Omega_{\Lambda}$.
We discuss this dependence on the basis of perspicuous figures. 
\end{abstract}
\newpage
\section{CBR Temperature Anisotropy.}

Although the Cosmic Background Radiation (CBR) was first discovered
over thirty years ago \cite{PW}, the detection of its temperature anisotropy
waited until 1992 when the Cosmic Background Explorer (COBE)
satellite provided its impressive experimental support\cite{smoot,ganga}
for the Big Bang model. In particular, the COBE results were consistent
with a scale-invariant spectrum of primordial scalar density
perturbations\cite{bar,sta,gupi,hawk} such as might be generated by quantum fluctuations
during an inflationary period.\cite{guth,lin,alb}

This discovery of temperature anisotropy in the CBR has 
inspired many further experiments which will be sensitive to
smaller angle anisotropies than the COBE satellite was (about $1^o$).
NASA has approved the flight of a satellite mission, the Microwave
Anisotropy Probe (MAP) in the year 2000 and ESA has agreed to
a more accurate later experiment called the Planck Surveyor.
The expected precision of these measurements implies that
the angular dependence of the temperature anisotropy will
be known sufficiently well that the location of the first
accoustic (Doppler) peak, and possibly subsequent ones, will
be resolved.

Although the hot big bang theory is supported by at least three
major triumphs: the expansion of the universe, the 
cosmic background radiation and the nucleosynthesis calculations,
it leaves unanswered several questions. The most important
unanswered questions are the horizon and flatness issues.

When the CBR last scattered, the age of the universe was
about 100,000 years compared to its present age
of some 10 billion years. As we shall see, the horizon size
at the recombination time subtends now an angle
of about $(1/208)$  of $\pi$ radians. On the celestial sphere
there are therefore approximately 40,000 causally
disconnected regions. Nevertheless, these different
regions have a uniform CBR temperature to an accuracy
of better than one part in $10^5$. This is the 
horizon problem.

The flatness problem may be understood from
the cosmological equation
\begin{equation}
\frac{k}{R^2} = (\Omega - 1) \frac{\dot{R}^2}{R^2}  \label{cos}
\end{equation}
Evaluating Eq.(\ref{cos}) at an arbitrary time $t$ and dividing by the
same relation at the present time $t=t_0$ and using $ R \sim \sqrt{t}
\sim T^{-1}$ gives
\begin{equation}
(\Omega-1) = 4H_0^2t^2\frac{T^2}{T_0^2} (\Omega_0 -1)   \label{omega}
\end{equation}
For high densities we write
\begin{equation}
\frac{\dot{R}^2}{R^2} = \frac{8 \pi G \rho}{3} = \frac{ 8 \pi Gga T^4}{6}
\end{equation}
where $a$ is the radiation constant and g is the effective number of degrees of freedom.
This leads to the relation between time and temperature, after
substituting the numerical values [$a=7.56\times10^{-9}erg m^{-3}K^{-4}; 
G/c^2 = 0.742\times 10^{-30}m/g; H_0=100h_0 km/s/Mpc = 3.25\times 10^{-18}h_0s^{-1}$]
\begin{equation}
t(seconds) = (2.42\times10^{-6})g^{-1/2}T^{-2}_{GeV}   \label{time}
\end{equation}
Combining Eq.(\ref{omega}) with Eq.(\ref{time}) leads to
\begin{equation}
(\Omega - 1) = 3.64\times10^{-21}h_0^2g^{-1}T_{GeV}^{-2}(\Omega_0 - 1)
\end{equation}
Given the proximity of $\Omega_0$ to unity, we then deduce that
$\Omega$ at, for example, $T=1MeV$ ($t\sim$ 1second) must be
equal to one within one part in $10^{14}$! Otherwise the resultant
cosmology will be incompatible with the present situation
of our universe. This extraordinary fine-tuning is the flatness problem.

The goal\cite{stein1,stein2,stein3,kam1,kam2,kam3,kam4} 
of the CBR experiments is to measure the temperature
autocorrelation function. The fractional temperature perturbation
as a function of the direction $\hat{{\bf n}}$ is expanded in
spherical harmonics
\begin{equation}
\frac{\Delta T(\hat{{\bf {n}}})}{T} = \sum_{lm} a_{(lm)}Y_{lm}(\hat{{\bf {n}}})
\end{equation}
and the statistical isotropy and homogeneity of the universe imply that
the coefficients have expectation values
\begin{equation}
<(a_{(lm)})^{*}a_{(l'm')}> = C_l\delta_{ll'}\delta_{mm'}
\end{equation}

The plot of $C_l$ versus $l$ is expected to reflect oscillations in
the baryon-photon fluid at the surface of last scatter. In particular,
the first Doppler peak should be at the postion $l_1 = \pi/\Delta\theta$
where $\Delta\theta$ is the angle now subtended by the horizon
at the time of the last scattering, namely the recombination time
corresponding to a red shift $z_t \sim 1,100$.

The horizon and flatness problems described above can both be solved
by the inflation scenario which has the further prediction
that $\Omega_0 = 1$ if the cosmological constant vanishes or
more generally that $\Omega_0 + \Omega_{\Lambda} = 1$ if
the cosmological constant does not vanish.

The question we address here is restricted to the question
of how much the value of $l_1$ alone - likely to be
accurately determined in the next few years -  will 
tell us about the values of the cosmic parameters $\Omega_0$
and $\Omega_{\Lambda}$?

In Section 2, the case $\Lambda=0$ is discussed. In Section 3,
there is the more general case; and finally in Section 4
there is discussion of the Figures derived.

\newpage
\bigskip
\bigskip

\section{The Special Case $\Lambda=0$, $0 < \Omega_0 < 1$}

When the cosmological constant vanishes, the Einstein-Friedmann
cosmological equations can be solved analytically (not the case,
in general, when $\Lambda\neq0$). So we shall begin by doing this
special case explicitly. It gives rise to the well-known
result that the position of the first Doppler peak (partial wave $l_1$)
expected
in the partial-wave analysis depending on the present 
matter-energy density $\Omega_0$ (for $\Lambda = 0$) 
according to $l_1 \sim 1/\sqrt{\Omega_0}$\cite{stein3,kam4}.
We shall show in the next section
how in the general case with $\Lambda\neq0$ there is a rather
serious "comic confusion" in disentangling the value of
$\Omega_0$ from the position $l_1$ of the first Doppler peak.

Let us use the metric:

\begin{equation}
ds^2 = dt^2 - R^2[d\Psi^2 + sinh^2\Psi d\theta^2 + sinh^2\Psi sin^2\theta d\phi^2]
\label{metric}
\end{equation} 
For a geodesic $ds^2=0$ and, in particular,
\begin{equation}
\frac{d\Psi}{dt} = \frac{1}{R}
\end{equation}
\noindent Einstein's equation reads
\begin{equation} 
\left( \frac{\dot{R}}{R} \right)^2 = \frac{8\pi}{3}G\rho + \frac{1}{R^2} \label{einstein} 
\end{equation}
where we take curvature $k=-1$. Let us define:
\begin{equation}
\Omega_0 = \frac{8 \pi G \rho_0}{3H_0^2};
\rho=\rho_0\left(\frac{R_0}{R}\right)^3; a=\Omega_0H_0^2R_0^3
\end{equation}
Then from Eq.(\ref{einstein}) we find that
\begin{equation}
\dot{R}^2R^2 = R^2 + aR \label{RR}
\end{equation}
and so it follows that
\begin{equation}
\frac{d\Psi}{dR} = \frac{d\Psi}{dt}\left(\frac{dR}{dt}\right)^{-1} = 
\frac{1}{\dot{R}R} = \frac{1}{\sqrt{R^2 + aR}}
\end{equation}
Since $\Psi_0 = 0$, the value at time $t$ can be computed from the integral
\begin{equation}
\Psi_t = \int_{R_t}^{R_0} \frac{dR}{\sqrt{(R + a/2)^2 - (a/2)^2}}  \label{integral}
\end{equation}
This can be performed easily with the substitution $R = \frac{1}{2}a(coshV - 1)$
to give the result:
\begin{equation}
\Psi_t = cosh^{-1} \left( \frac{2R_0}{a} + 1\right) - 
cosh^{-1} \left( \frac{2R_t}{a} + 1\right) \label{Psi}
\end{equation}
>From Eq.(\ref{einstein}) evaluated at $t=t_0$ we see that
\begin{equation}
\frac{1}{a} = \frac{1 - \Omega_0}{R_0 \Omega_0}
\end{equation}
and so, using $sinh(cosh^{-1}x)=\sqrt{x^2-1}$ in Eq.(\ref{Psi}) gives now
\begin{equation}
sinh\Psi_t = \left( \sqrt{\left(\frac{2(1-\Omega_0)}{\Omega_0} + 1\right)^2 - 1} \right) \left( \frac{2R_t}{a} + 1 \right) -
\left( \sqrt{\left(\frac{2(1-\Omega_0)R_t}{\Omega_0R_0} + 1\right)^2 - 1} \right) \left( \frac{2R_0}{a} + 1 \right) 
\label{sinh}
\end{equation}
The position of the first Doppler peak depends on the angle subtended
by the horizon size at the time $t$ equal to the recombination time.
This corresponds to the distance $(H_t)^{-1}$. According to the metric
of Eq.(\ref{metric}) the angle subtended is
\begin{equation}
\Delta \theta = \frac{1}{H_t R_t sinh \Psi_t}     \label{theta}
\end{equation}
and the position of the first Doppler peak corresponds to the partial
wave $l_1$ given by
\begin{equation}
l_1 = \frac{\pi}{\Delta \theta} = \pi H_t R_t sinh \Psi_t    \label{l}
\end{equation}
Now the red-shift at recombination is about $z_t=1100 \simeq (R_0/R_t) \gg 1$
so we may approximate in Eq.(\ref{sinh}) to find
\begin{equation}
sinh\Psi_t \simeq \frac{2 \sqrt{1 - \Omega_0}}{\Omega_0}
\end{equation}
Using $H_t^2 = 8\pi G \rho/3 + 1/R^2 \simeq \Omega_0H_0^2(R_0/R_t)^3$
gives
\begin{equation}
l_1(\Lambda = 0) = \frac{2 \pi}{\sqrt{\Omega_0}} z_t^{1/2}
\end{equation}
In particular, if $\Omega_0 = 1$ and $\Lambda = 0$, one has $l_1 \simeq 208.4$.
If $l_1$ does have this value empirically it will favor this
simplest choice, although as we shall see in the following subsection
even here the conclusion has ambiguities.

In Fig. 1 we plot $l_1$ versus $\Omega_0$ for the particular case of
$\Omega_{\Lambda} = 0$.

\newpage

\section{The General Case: $0 \leq \Omega_0 < 2; 0 \leq \Omega_{\Lambda} < 1$}

For the general case of $0 \leq \Omega_{\Lambda} < 2; 0 < \Omega_0 < 1$
we use the more general Einstein cosmological equation:
\begin{equation}
\dot{R}^2R^2 = -k R^2 + aR + \Lambda R^4 /3   \label{geneinst}
\end{equation}
where $a = \Omega_0 H_0^2 R_0^3$. We define
\begin{equation}
\Omega_0 = \frac{8 \pi G \rho_0}{3 H_0^2}; \Omega_{\Lambda} = 
\frac{\Lambda}{3 H_0^2}; \Omega_C = \frac{-k}{H_0^2 R_0^2}
\end{equation}
Substituting $R = R_0 r$ and $w = 1/r$ now gives rise to the integral\cite{int} for $\Psi_t$
\begin{equation}
\Psi_t = \sqrt{\Omega_C} \int_1^{\infty} \frac{dw}{\sqrt{\Omega_{\Lambda} + \Omega_C w^2 + \Omega_0 w^3}}
\end{equation}
in which $\Omega_{\Lambda} + \Omega_{C} + \Omega _0 = 1$.

Consider first the case of an open universe $\Omega_C > 0$. Then
\begin{equation}
l_1 = \pi H_t R_t sinh\Psi_t
\end{equation}

We know that 
\begin{eqnarray}
H_t^2 & = & \left( \frac{\dot{R}_t}{R_t} 
\right)^2 = \frac{8 \pi G \rho}{3} + \frac{\Lambda}{3} + \frac{1}{R_t^2} \nonumber \\
& = & H_0^2 \left[ \Omega_0 \left( \frac{R_0}{R_t}\right)^3 + 
\Omega_{\Lambda} + \left( \frac{R_0}{R_t}\right)^2 \Omega_C \right]
\end{eqnarray}
Since $R_0 \gg R_t$ we may approximate:
\begin{equation}
H_t \simeq \left( \frac{R_0}{R_t} \right)^{3/2} H_0 \sqrt{\Omega_0}
\end{equation}
and hence
\begin{equation}
H_tR_t = \left( \frac{R_0}{R_t} \right)^{1/2} \sqrt{\frac{\Omega_0}{\Omega_C}}
\end{equation}
It follows that for this case $\Omega_C > 0$ that
\begin{equation}
\l_1 = \pi \sqrt{\frac{\Omega_0}{\Omega_C}} 
\left(\frac{R_0}{R_t}\right)^{1/2} sinh \left( \sqrt{\Omega_C} \int_1^{\infty} 
\frac{dw}{\sqrt{\Omega_{\Lambda} + \Omega_C w^2 + \Omega_0 w^3}} \right)
\label{l1c+}
\end{equation}
For the case $\Omega_C < 0 (k = +1)$ we simply replace $sinh$ by $sin$ in Eq. (\ref{l1c+}).
Finally, for the special case $\Omega_C = 0$, the generalized flat case favored by
inflationary cosmologies, Eq.(\ref{l1c+}) simplifies to:
\begin{equation}
\l_1 = \pi \sqrt{\Omega_0} 
\left(\frac{R_0}{R_t}\right)^{1/2}  \int_1^{\infty} 
\frac{dw}{\sqrt{\Omega_{\Lambda} + \Omega_0 w^3}}
\end{equation}
In Fig. 2 we plot the value of $l_1$ versus $\Omega_0$ for
the case $\Omega_C = 0$ (flat spacetime). The contrast with
Fig 1 is clear: whereas $l_1$ increases with decreasing $\Omega_0$
when $\Omega_{\Lambda}=0$ (Fig. 1) the opposite behaviour occurs when
we constrain $\Omega_{\Lambda} = 1 - \Omega_0$ (fig.2).

With $\Omega_0$ and $\Omega_{\Lambda}$ unrestricted there are more
general results. In Fig. 3, we display iso-l lines on a 
$\Omega_0 - \Omega_{\Lambda}$ plot.
The iso-l lines are (from right to left) for the values
$l_1 = 150, 160, 170, 180, 190, 200, 210, 220, 230, 240, 250, 260, 270$ respectively.
One can see that from the {\it position}($l_1$)  only of the first Doppler peak
there remains a serious ambiguity of interpretation without further information.

In Fig, 4, there is a three dimensional rendition of the
value of $l_1$ versus the two variables $\Omega_0$ and $\Omega_{\Lambda}$.

\newpage

\section{Discussion of Cosmic Parameter Ambiguities.}

Let us now turn to an interpretation of our Figures, from
the point of view of determining the cosmic parameters. 

In the case where $\Lambda = \Omega_{\Lambda}=0$, Fig.1. is
sufficient. In this case, there is the well-known dependence\cite{stein3,kam4}
$l_1 = (208.4)/\sqrt{\Omega_0}$ illustrated in Fig.1.
It would be straightforward to determine $\Omega_0$ with
an accuracy of a few percent from the upcoming measurements.

Of course there is a strong theoretical prejudice towards
$\Lambda=0$. But no underlying symmetry principle is yet
known. If $\Omega_{\Lambda} \neq 0$, one knows that it
is not bigger than order one; this is very many orders
of magnitude smaller than expected\cite{WN} from the vacuum energy
arising in spontaneous breaking of symmetries such as the electroweak
group $SU(2) \times U(1)$.

Nevertheless, recent observations of high redshift Type 1a supernovae
have led to the suggestion of an {\it increasing} Hubble parameter
\cite{perl1,perl2}. An interpretation of this is that 
the cosmological constant is non-zero, possibly
$\Omega_{\Lambda} \simeq 0.7$ but is still consistent
with $\Omega_0 = 1 - \Omega_{\Lambda}$. It should be added
that these conclusions are quite controversial and await
further verification. But these results are enough to motivate
a full consideration of non-zero values of $\Omega_{\Lambda}$.

Thus we come to Fig. 2 which depicts the $\Omega_0$ dependence
of $l_1$ when $\Omega_0 + \Omega_{\Lambda} = 1$ is held fixed
as in a generalized flat cosmology that could arise from inflation.
We notice that here $l_1$ {\it decreases} as $\Omega_0$
decreases from $\Omega_0 = 1$, the opposite behaviour
to Fig. 1. Thus even the sign of the shift of $l_1$ from $l_1 =208.4$
depends on the size of $\Lambda$.

It is therefore of interest to find what are the contours
of constant $l_1$ in the $\Omega_0 - \Omega_{\Lambda}$ 
plane. These iso-l lines are shown in Fig. 3 for
$l_1 = 150,....,270$ in increments $\Delta l_1 = 10$.
If we focus on the $l_1 = 210$ contour (the seventh contour from
the left in Fig. 3) as an example,
we see that while this passes close to the
$\Omega_0 = 1, \Lambda = 0$ point it also tracks out
a line naturally between those shown in Figs. 1 and 2 (actually
somewhat closer to the latter than the former).

Finally, Fig. 4 gives a three-dimensional rendition which
includes Figures 1 to 3 as special cases and provides a visualisation
of the full functional dependence of $l_1(\Omega_0, \Omega_{\Lambda})$.

Our main conclusion is that the position $l_1$ of the first
Doppler peak will define the correct contour in our
iso-l plot, Fig. 3. More information will be necessary
to determine $\Omega_0$ and the validity of inflation.

\bigskip
\bigskip
\bigskip
\bigskip
\bigskip
\bigskip
\bigskip
\bigskip
\bigskip
\bigskip

We thank Eric Carlson of Wake Forest University 
for useful discussions, and Masayasu Harada for help.
This work was supported in part by the
US Department of energy under Grant No. 
DE-FG05-85ER-40219.

\bigskip
\bigskip
\bigskip
\bigskip
\bigskip
\bigskip
\bigskip
\bigskip
{\bf Note Added.}

After completing this paper, three very recent papers 
having some overlap with our work were brought to our attention:

M. White. {\it astro-ph/9802295}; M. Tegmark, D.J. Eisenstein, W. Hu and R.G. Kron. {\it astro-ph/9805117};
C.H. Lineweaver. {\it astro-ph/9805326}.

\vspace{20mm}

{\bf Figure Captions.}

\noindent Fig 1. Plot of $l_1$  vs. $\Omega_0$ for $\Omega_{\Lambda} = 0$.

\noindent Fig. 2. $l_1$ vs. $\Omega_0$ for the case $\Omega_{\Lambda} = 1 - \Omega_0$.

\noindent Fig. 3. Iso-l lines on $\Omega_0 - \Omega_{\Lambda}$ plot,
for (from right to left)\\
\indent  $l_1 = 150$ through $270$ in increments $\Delta l = 10$.
Horizontal = $\Omega_0$, Vertical = $\Omega_{\Lambda}$.

\noindent Fig. 4. Three-dimensional plot of $l_1$ against $\Omega_0$ and
$\Omega_{\Lambda}$. Front = $\Omega_{\Lambda}$, Right = $\Omega_0$.

\newpage
\begin{figure}
\begin{center}
\epsfxsize=6.0in
\ \epsfbox{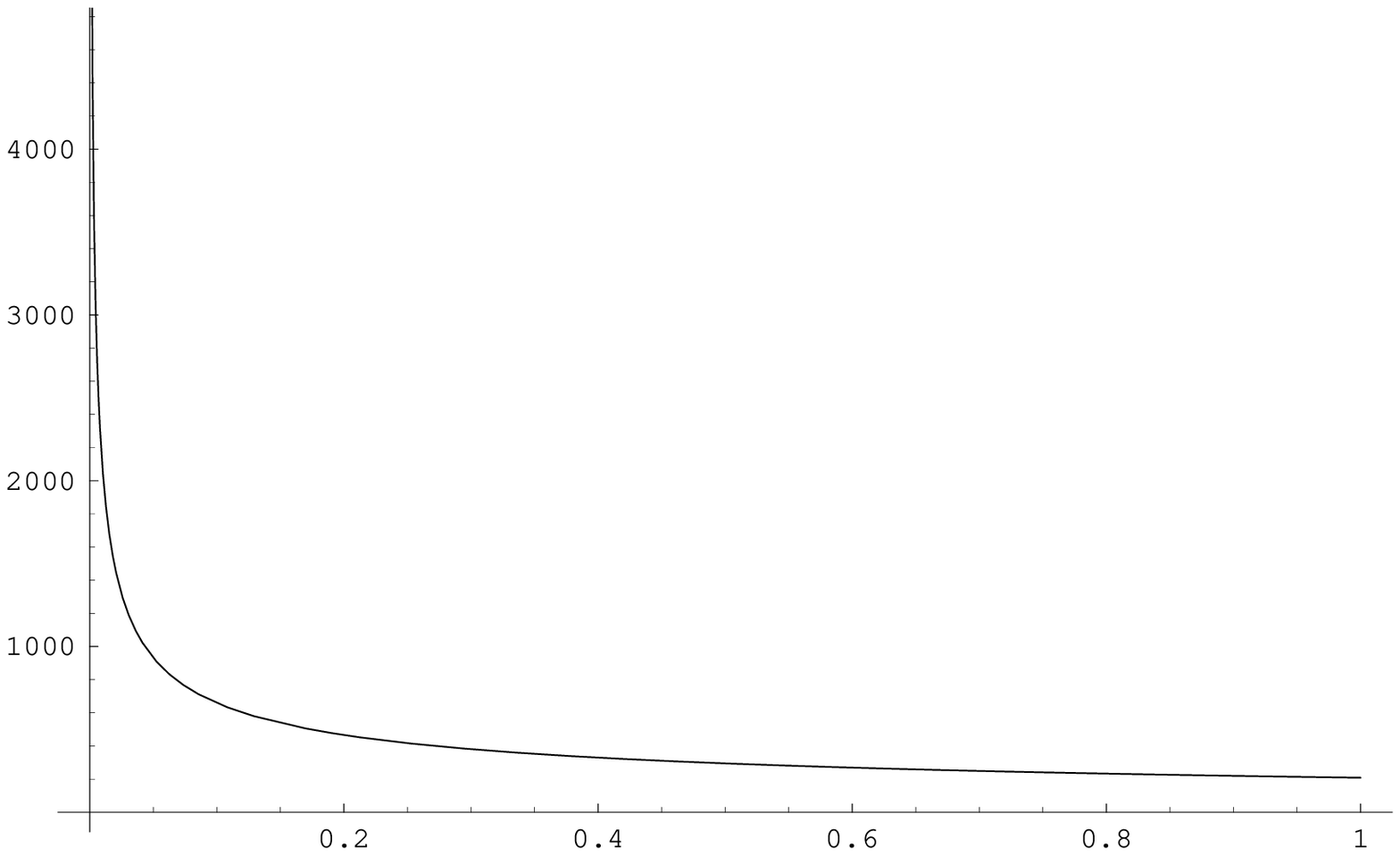}
\end{center}
\end{figure}

\begin{center}
Figure 1.
\end{center}

\newpage
\begin{figure}
\begin{center}
\epsfxsize=6.0in
\ \epsfbox{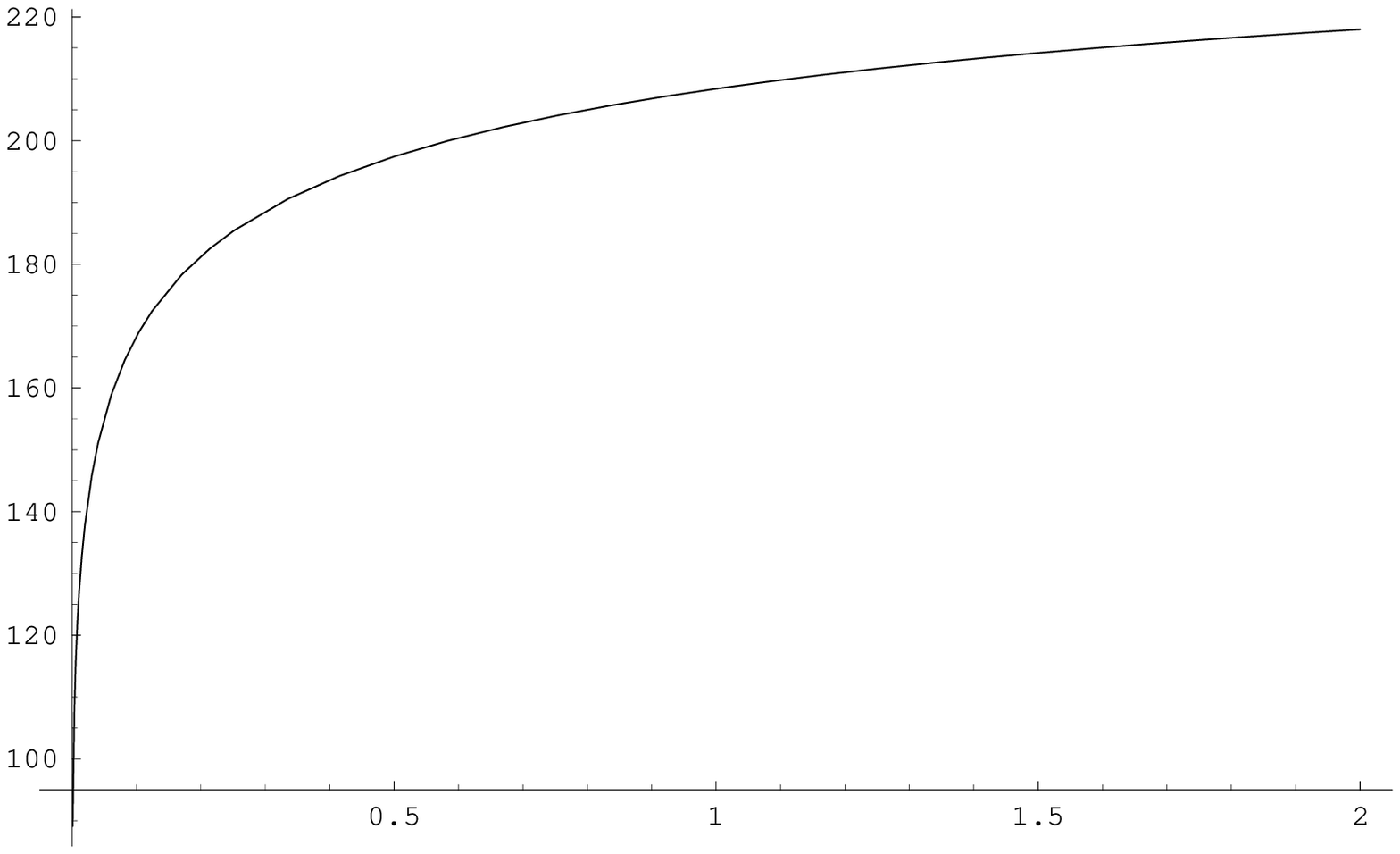}
\end{center}
\end{figure}

\begin{center}
Figure 2.
\end{center}

\newpage
\begin{figure}
\begin{center}
\epsfxsize=6.0in
\ \epsfbox{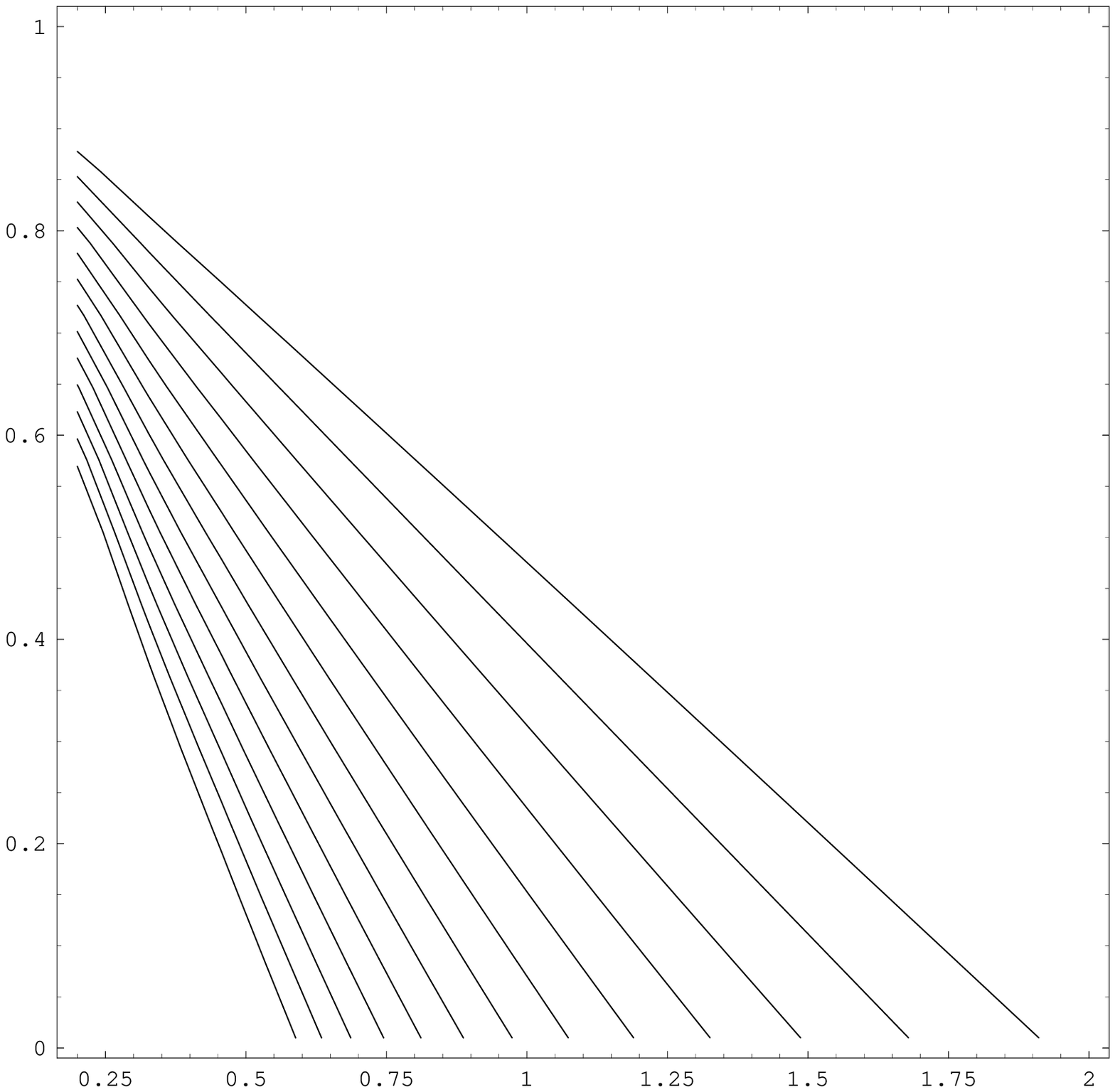}
\end{center}
\end{figure}

\begin{center}
Figure 3.
\end{center}

\newpage
\begin{figure}
\begin{center}
\epsfxsize=6.0in
\ \epsfbox{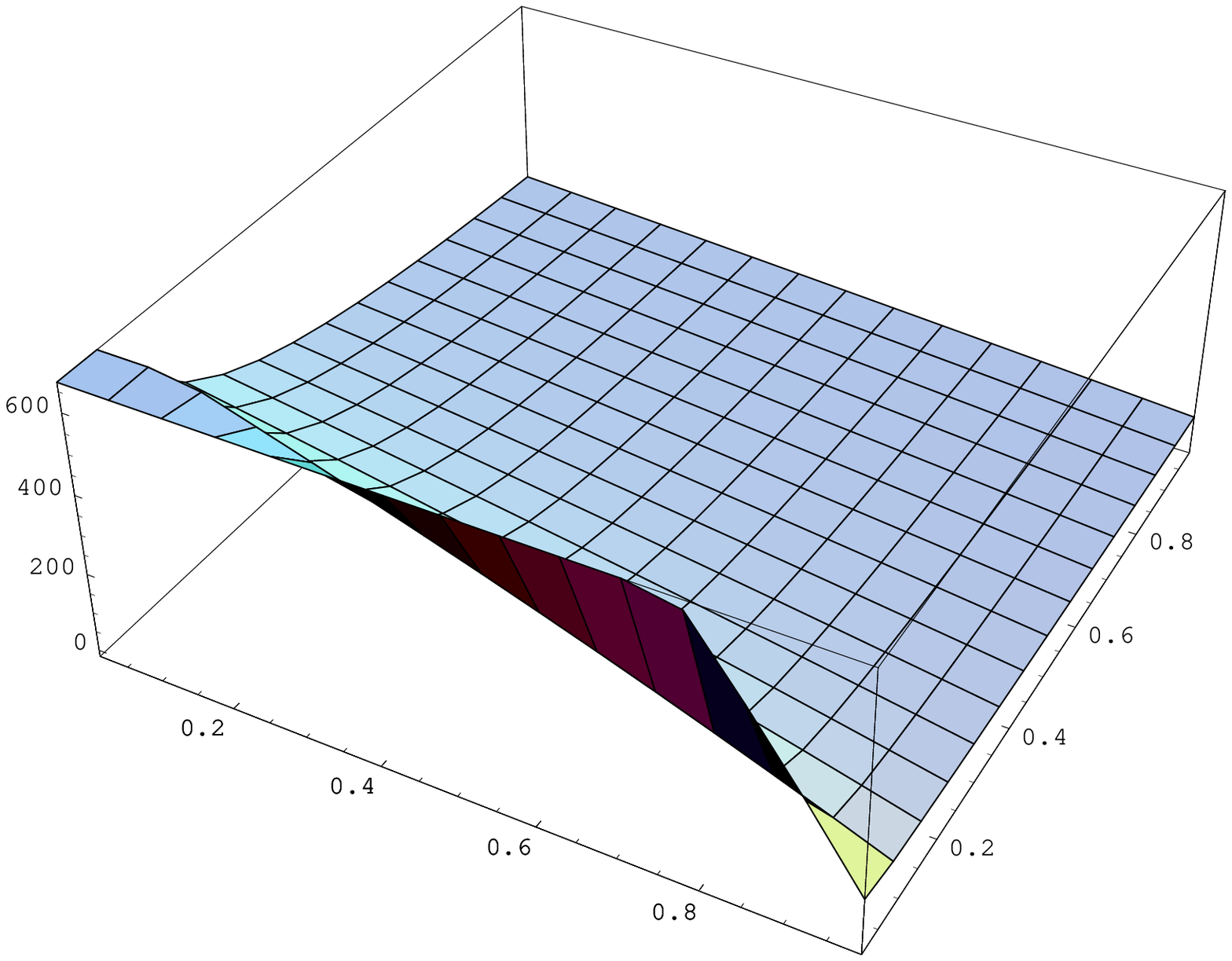}
\end{center}
\end{figure}

\begin{center}
Figure 4.
\end{center}


\begin{thebibliography}{99}
\bibitem{int}
We note the similarity to the formula for the age of the universe:
\begin{equation}
t_0 = \frac{1}{H_0}\int_1^{\infty}\frac{dw}{w\sqrt{\Omega_{\Lambda} 
+ \Omega_C w^2 + \Omega_0 w^3}}    \nonumber
\end{equation}
which is also an elliptic integral that cannot be done analytically.
\end{thebibliography}
\end{document}